# Enhancing Innate and Adaptive Immune Systems by Cold Atmospheric Plasma (CAP) and Its Antitumor Immunity

Fengdong Cheng[1,†], Dayun Yan[2,†], Jie Chen[1], Zi Wang[1], Alex Horkowitz[2], Michael Keidar[2]* and Eduardo M. Sotomayor[1]*.

[1]George Washington Cancer Center, George Washington University, Washington DC, USA
[2]School of Engineering and Applied Science, George Washington University, Washington DC, USA

*Corresponding Authors Contact Information:
Eduardo M. Sotomayor, MD
George Washington Cancer Center
Science & Engineering Hall, Suite 8000
800 22nd Street, NW, Washington, DC 20052
Phone: (202) 994-0329
Email: eduardo.marcona57@gmail.com

or

Michael Keidar, Ph.D.
Mechanical and Aerospace Engineering
School of Engineering and Applied Science
The George Washington University
Science & Engineering Hall 3550
800 22nd Street, NW, Washington, DC 20052
Phone: 202-994-6929
email: keidar@gwu.edu

[†] These authors contributed equally to this work.

**Abstract:** Cold atmospheric plasma (CAP) is a near room temperature ionized gas, generated under non-equilibrium discharge conditions. Here we show that a short exposure of rat peritoneal exudate macrophages and T-cells to CAP *in vitro*, triggered an inflammatory phenotype leading to better antigen-presenting and effector cell function respectively. Different from previous studies mainly using immortalized cell lines, both macrophage and T-cells in this study were primary cells isolated from mice. Furthermore, *ex-vivo* exposure of T-cells to CAP, followed by their adoptive transfer into tumor-bearing mice resulted in a strong antitumor effect *in vivo*. Mechanistically, CAP seems to disrupt tolerogenic pathways leading to enhanced production of pro-inflammatory cytokines while limiting the production of anti-inflammatory cytokines and the expression of inhibitory molecules such as programmed death-ligand 1 (PD-L1). CAP represents therefore a novel, non-toxic and easy to deliver technology to augment the function of immune cells and enhance antitumor responses when used as a component of T-cell adoptive immunotherapies strategies or, potentially in combination with other cancer immunotherapeutic approaches.

**Introduction:** In recent years, the field of cancer immunotherapy has revolutionized cancer care [1,2]. Treatment of patients with advanced solid malignancies or hematologic cancers with either checkpoint blockade antibodies or with genetically engineered T-cells (CAR T-cells) has resulted in impressive clinical responses and potential cures [3]. However, not all cancer patients benefit with these therapies and as such, novel technologies able of enhancing antitumor immune responses and are easy to deliver, must be identified to further expand the clinical efficacy of the current generation of cancer immunotherapies.

Cold atmospheric plasma (CAP) is a near room temperature ionized gas generated under non-equilibrium discharge conditions [4,5]. CAP is composed of numerous ionized products including neutral particles such as atoms, molecules, and charged particles such as electrons, ions, and long/short-lived reactive species such as reactive oxygen/nitrogen species (ROS/RNS) [6]. In many references, CAP is also named cold plasma, non-thermal plasma, gas plasma, or physical plasma [7–9]. CAP's multimodal chemical nature makes it a flexible, controllable, and self-adaptive modality in many medical applications, including microorganism sterilization, wound healing, and cancer therapy [10,11]. CAP has shown impressive visions as a novel anti-cancer tool by selectively killing cancer cell lines *in vitro* and diminishing the xenografted subcutaneous tumor in a non-invasive approach [12,13]. ROS/RNS has been regarded as the main player to cause a CAP treatment's main biological effects, such as cell death, cycle arrest, and autophagy [14].

Rounding up these unique CAP's properties, emerging data indicate that CAP is a potential novel modality to activate the immune response and cause immunogenic cancer cell death [15]. A uniform nanosecond pulsed CAP treatment activated macrophages can improve an artificial wound's healing efficacy [16]. CAP triggered cancer cells to emit damage-associated molecular pattern signals to stimulate local immune cells [17]. For example, the increased calreticulin level on the surface of CAP-treated cancer cells has been observed in several cases [18]. The secretion of ATP has also been activated in cancer cells after CAP treatment, particularly in combination with other immunotherapies [19]. In another animal study, the production of inflammatory cytokine (IFN-γ) from splenocytes was also increased after a CAP treatment on melanoma model on mouse legs [20]. Additionally, the micromolar level of $H_2O_2$ secretion from diverse cancer cell lines has been observed immediately after a CAP treatment lasting just 1 min [21,22]. The cell-based secretion will continue even CAP treatment has been stopped [23]. $H_2O_2$ is a second messenger in lymphocyte activation [24]. Micromole levels of $H_2O_2$ rapidly induce the expression of interleukin-2 (IL-2) [25].

However, little is still known about the effects of CAP upon cells of the innate and adaptive immune system, particularly T-cells, and which could be the best strategy to efficiently trigger CAP-mediated antitumor immune responses *in vivo*. Here we show that *in vitro* exposure of primary macrophages or T-cells to CAP, not only is not toxic but triggers instead of an inflammatory phenotype that leads to a better antigen-presenting cell and effector cell function, respectively. Besides, *ex-vivo* exposure of T-cells to CAP followed by their adoptive transfer into tumor-bearing mice resulted in a stronger antitumor effect *in vivo*. CAP represents, therefore, a novel, non-toxic, and easy to deliver technology to augment the function of cells of the immune system and enhance antitumor responses when used as a component of T-cell adoptive immunotherapies [26].

**Material and methods.**

**CAP jet and treatment:** An illustration of the CAP jet source used in our experiments is shown in **Fig. 1A.** Briefly, the discharge (3 kV (rms)) of helium gas (Roberts Oxygen, grade 4.5, size 300) was initiated between a central anode and an annular cathode with a pulse frequency of 12.5 kHz. If discharge did not occur, CAP jet would not be formed. In this case, only helium gas flow would affect targets (cells or tissues). When discharge occurred, the ionized gas flowed out through a glass tube (1.5 lpm). The outlet ionized gas formed a stable violet gas jet, composed of long/short-lived reactive species and other highly reactive components. These reactive species would cause a biological impact on samples. The jet tip's temperature was less than 40°C. Displayed in **Fig. 1B** is a schematic illustration of this study's basic research strategy. As an example, CAP treatment was performed on the purified immune cells. Peritoneal elicited macrophages (PEM) or T-cells were isolated from the spleen of C57BL/6 mice, were seeded in multi-wells plates. CAP treatment *in vitro* was performed on them (0-120 seconds), followed by biological assays including cytokine production, surface markers, and antigen-presentation.

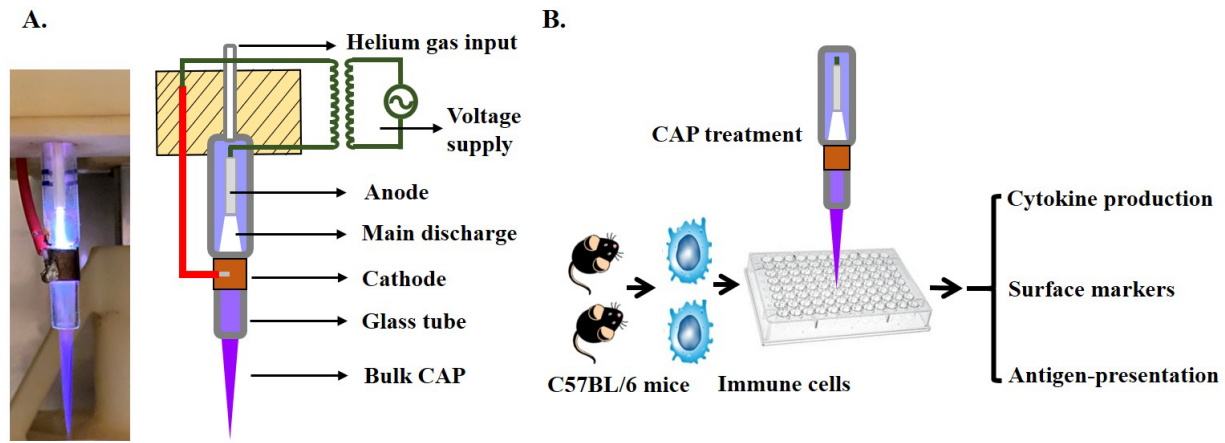

Figure 1. CAP device and strategy for treatment of immune cells *in vitro*. (A) Schematic illustration of the CAP jet source used to treat immune cells *in vitro*. (B) Research strategy.

**Mice and cell line:** All mice were housed in pathogen-free conditions in the same designated room of the animal facility at George Washington University. All animal studies were performed according to the guideline of approved protocols by the Institutional Animal Care and Use Committee (IACUC) at George Washington University. EL-4 murine thymoma was purchased from American Type Culture Collection (ATCC). EL-4 OVA is a cell line transfected to express Ovalbumin in their surface. Both cell lines were cultured and maintained in Dulbecco's modified Eagle's medium (DMEM, Corning) supplemented with 4 mM L-glutamine, 4500 mg/L glucose, 1mM sodium pyruvate, 1500 mg/L sodium bicarbonate, 10% horse serum (ATCC), 100 U/mL penicillin and 100 μg/mL streptomycin. Cell line was grown under humidified conditions at 5% $CO_2$ and 37ºC.

***In Vivo* tumor model:** For *in vivo* tumor studies, $0.25 \times 10^6$ EL-4 cells were injected subcutaneously into the right flank of female mice that were 6-8 weeks old. Tumor volume was measured starting around Day 8 after injection or at the time when tumors were palpable.

**Splenic and lymph node cells isolation:** Mouse spleens or lymph nodes were harvested under sterile conditions. Total splenocytes or lymphoid cells were obtained by mashing mouse spleens or lymph nodes through a 70 μm sterile cell strainer (Fisher Scientific). The cells were collected and red blood cells were lysed by incubating with ACK lysis buffer (Gibco).

**Flow cytometry analysis of cell surface marker:** For flow cytometric analysis, cells were stained with fluorochrome-labeled monoclonal antibodies (mAbs; anti-PD-L1, anti-TIM) for 20 min at 4ºC and washed for 2 times with FACS buffer. The vitality dye 4', 6-diamidino-2-phenylindole (DAPI, 50 ng/mL, Sigma) was added to cells before analysis. Data was acquired on FACS Celesta flow cytometer (BD Biosciences), at least 10,000 events, of the smallest population of interest, subsequently analyzed using FlowJo software 10.2.

**Statistics**

Statistical analyses were performed using student's t-test on Microsoft Excel. Data expressed as the mean ± SD. Probability values of $p \leq 0.05$ were considered significant.

**Results.**

**CAP enhances the macrophages function *in vitro* by triggering an inflammatory phenotype.**

Initial studies showed that the viability of rat peritoneal exudate macrophages (PEM) cultured *in vitro* with supplemented media was not affected when they were exposed to CAP for increasing periods of time up to 120 seconds. Instead, such an exposure triggered an inflammatory phenotype. First, the exposure of PEM to CAP (120 seconds), resulted in decreased expression of the immunosuppressive molecule PD-L1 **(Fig. 2A)**. In addition, PEM exposed to CAP for 15 or 30 seconds produced low levels of the anti-inflammatory cytokine, IL-10 and high levels of the pro-inflammatory cytokines IL-6 and IL-12 **(Fig. 2B, 2C).** Of note, this response was elicited in the absence of lipopolysaccharide (LPS), which is routinely used to trigger cytokine production by PEM. This was an unexpected finding, given that the levels of pro-inflammatory cytokines triggered by CAP were similar in magnitude to those elicited in LPS-treated PEM [26,27]. Gas treatment means the treatment was performed via only using helium gas flow to affect PEM. In this case, there was no discharge in helium. Just helium flow did not exert any impact on cells or tissues.

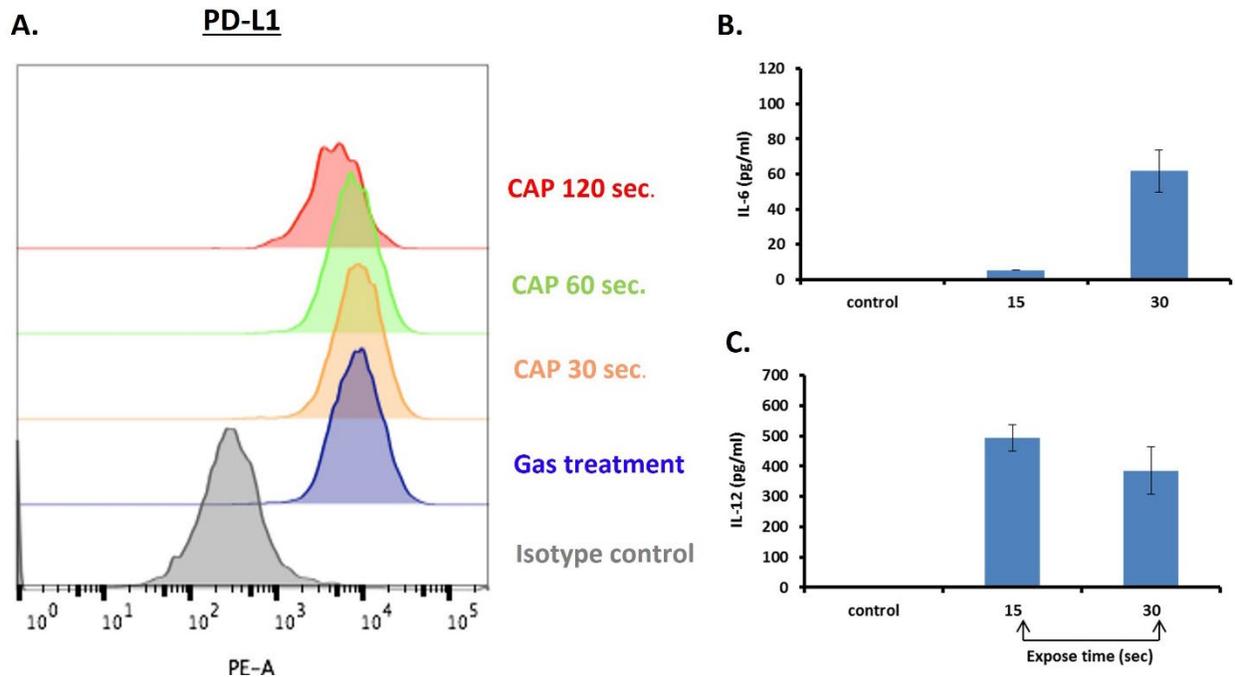

Figure 2. Phenotypic and Functional changes in PEM exposed to CAP. **(A)** PEM were isolated from C57BL/6 mice and exposed *in vitro* to CAP as indicated. After 24 hours, macrophages were harvested and the expression of PD-L1 was determined by using an anti-PD-L1 antibody and measured by flow cytometry. The data was analyzed by FlowJo software. **(B,C)** Supernatants were also collect and production of IL-12 and IL-6 were determined by ELISA. (n=3).

**PEM exposed to CAP are better activators of antigen-specific CD4+ T-cells.**

Given the pro-inflammatory phenotype displayed by CAP-treated PEM, we hypothesized that these cells would be better activators of antigen-specific T-cells *in vitro*. Briefly, PEMs were isolated from C57BL/6 mice, cultured *in vitro* with supplemented medium and exposed to CAP for 15 or 30 seconds (**Fig. 3A**). After 24 hours, purified CD4+ T-cells isolated from the spleen of OT-II mice were added to the PEM cultures. OVA$_{323-339}$ peptide has been used to investigate class II MHC-peptide binding and T-cells activation. In control group, OVA peptide has not been provided in PEM-T cell culture. In experimental groups, OVA peptide was added to the PEM-T cell cultures for 48 hours and supernatants were collected for assessment of cytokine production **(Fig. 3A)**. Antigen-specific CD4+ T-cells encountering cognate antigen in CAP-treated PEM produced higher levels of IL-2 and IFN-γ relative to T-cells cultured with OVA$_{323-339}$ peptide (not exposed to CAP) **(Fig. 3B, 3C).** Taken together, short exposure of PEM to CAP *in vitro*, does not affect their viability, triggering instead a pro-inflammatory phenotype that leads to a better antigen-presentation to antigen-specific T-cells.

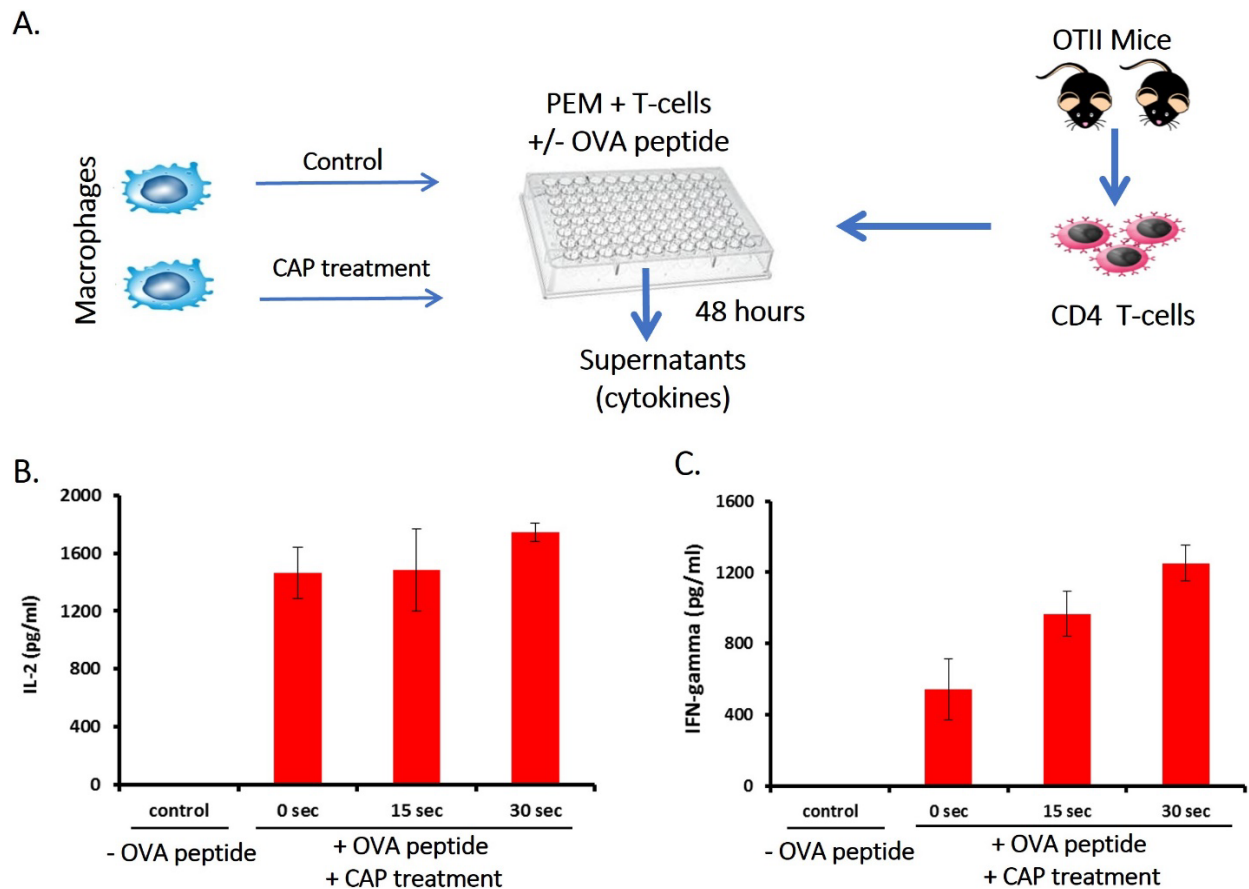

**Figure 3. PEM exposed to CAP are better activators of antigen-specific CD4+ T-cells**. **(A)** Illustration of antigen-presentation experiments. PEMs were isolated from C57BL/6 mice and exposed to CAP for different periods of time as indicated. Control: without OVA peptide and CAP treatment. OVA peptides were added in 0, 15, and 30 seconds of CAP treatment. After 24 hours, PEM were washed with warm RPMI. Then, purified CD4+ T-cells isolated from the spleen of OT-II mice were added to the PEM in presence -or not- of OVA$_{323-339}$ peptide for 48h. Supernatants were then collected. **(B,C)** Production of IFN-γ or IL-2 by antigen-specific T-cells was determined by ELISA. (n=3).

**Increased cytokine production by T-cells *in vitro*.**

Next, we assessed the CAP's effect upon purified T-cells *in vitro*. T-cells were isolated from the lymph nodes of C57BL/6 mice and were exposed to CAP (experimental groups) for 60 or 120 seconds followed by the stimulation with anti-CD3+anti-CD28+ antibodies (a well-known combination to induce cytokine production by T-cells) **(Fig. 4A).** Unlike PEM that produced inflammatory cytokines just by being exposed to CAP, T-cells did not product effector cytokines in the absence of simulation with anti-CD3anti-CD28+ regardless of with or without CAP treatment (**Fig. 4B** and **4C**). However, T-cells exposed to CAP and then stimulated with anti-CD3+anti-CD28+ produced higher levels of IL-2 and IFN-γ as compared to control T-cells (just exposure to helium flow without discharge and subsequently treated with anti-CD3+anti-CD28+) (**Fig. 4B** and **4C**). Therefore, a short exposure of T-cells to CAP *in vitro*, enhanced their production of cytokines that are critical in orchestrating effective antitumor immune responses.

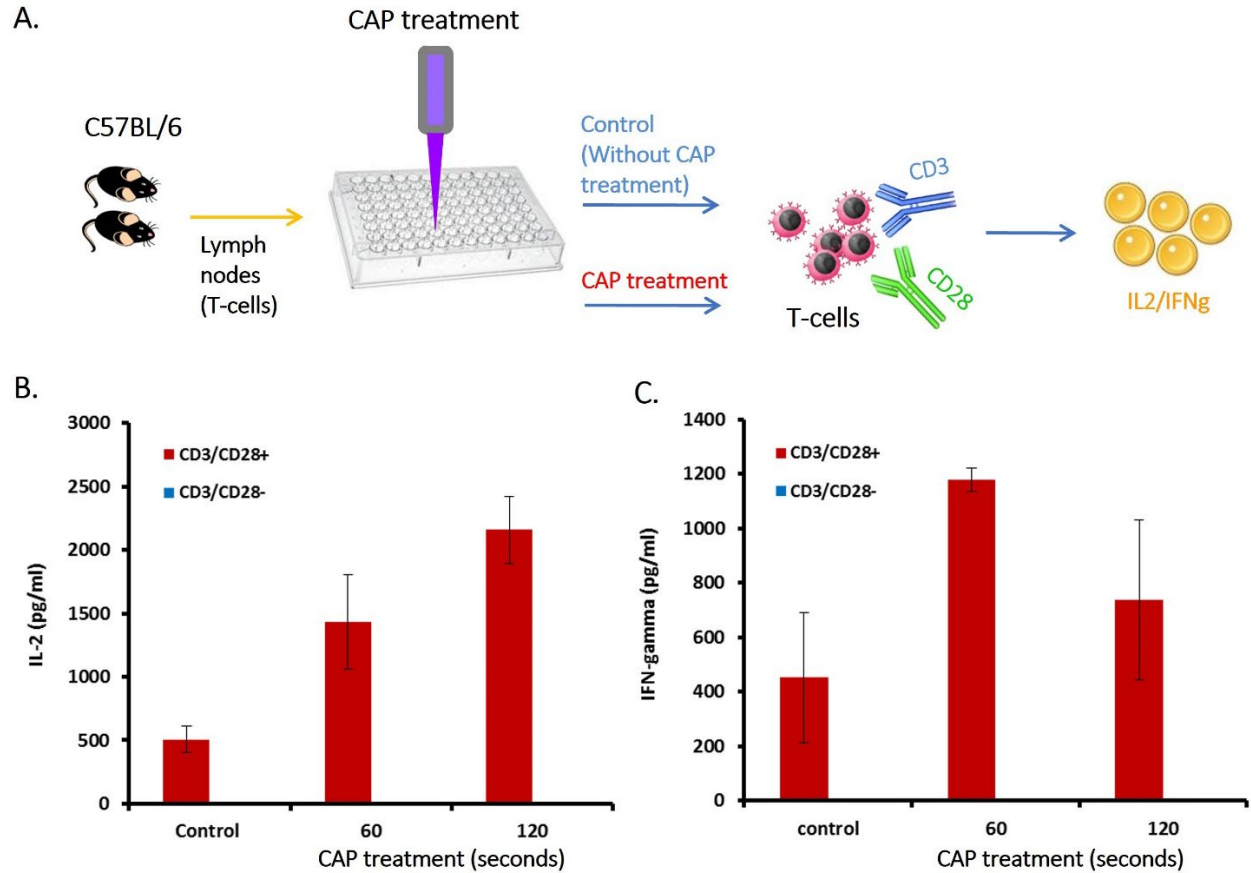

**Figure 4. Increased cytokine production by purified T-cells exposed to CAP. (A)** Experimental design for *in vitro* treatment of T-cells with CAP. T cells were purified from the lymph nodes of C57BL/6 mice and exposed to CAP for 60 or 120 seconds followed by stimulation with anti-CD3+anti-CD28+ antibodies. 48 hours later, supernatants were collected for assessment of cytokine production. **(B,C)** IL-2 or IFN-γ production were determined by ELISA. (n=3).

To investigate whether the microenvironment in lymph nodes can cause different impact on T-cells after the CAP treatment, we performed *ex-vivo* CAP treatment on lymph nodes. General illustration was shown in **Fig. 5A**. To gain insights into the effects of CAP upon T-cell responses *in vivo*, lymph nodes were isolated from OT-I mice (to assess antigen-specific CD8+T-cell responses) and plated in 6-wells plate filled with culture media and treated *ex-vivo* with either CAP or helium control for 60 seconds or, left untreated. Lymph nodes were then disaggregated, and isolated T-cells were cocultured in 96-wells plate with macrophages (isolated from C57BL/6 mice) in the presence of OVA peptide $_{257-264}$ (for CD8 T-cells cultures) for 48 hours. Supernatants were collected and cytokines IL-2 or IFN-γ production were determined by ELISA. As shown in **Fig. 5B** and **5c**, antigen-specific CD8+T-cells isolated from lymph nodes of OT-I mice that were *ex vivo* exposed to CAP, produced higher levels of IL-2 and IFN-γ in response to cognate peptide as compared to CD8+ T-cells from lymph nodes exposed to helium gas treatment or CD8+ T-cells from untreated lymph nodes.

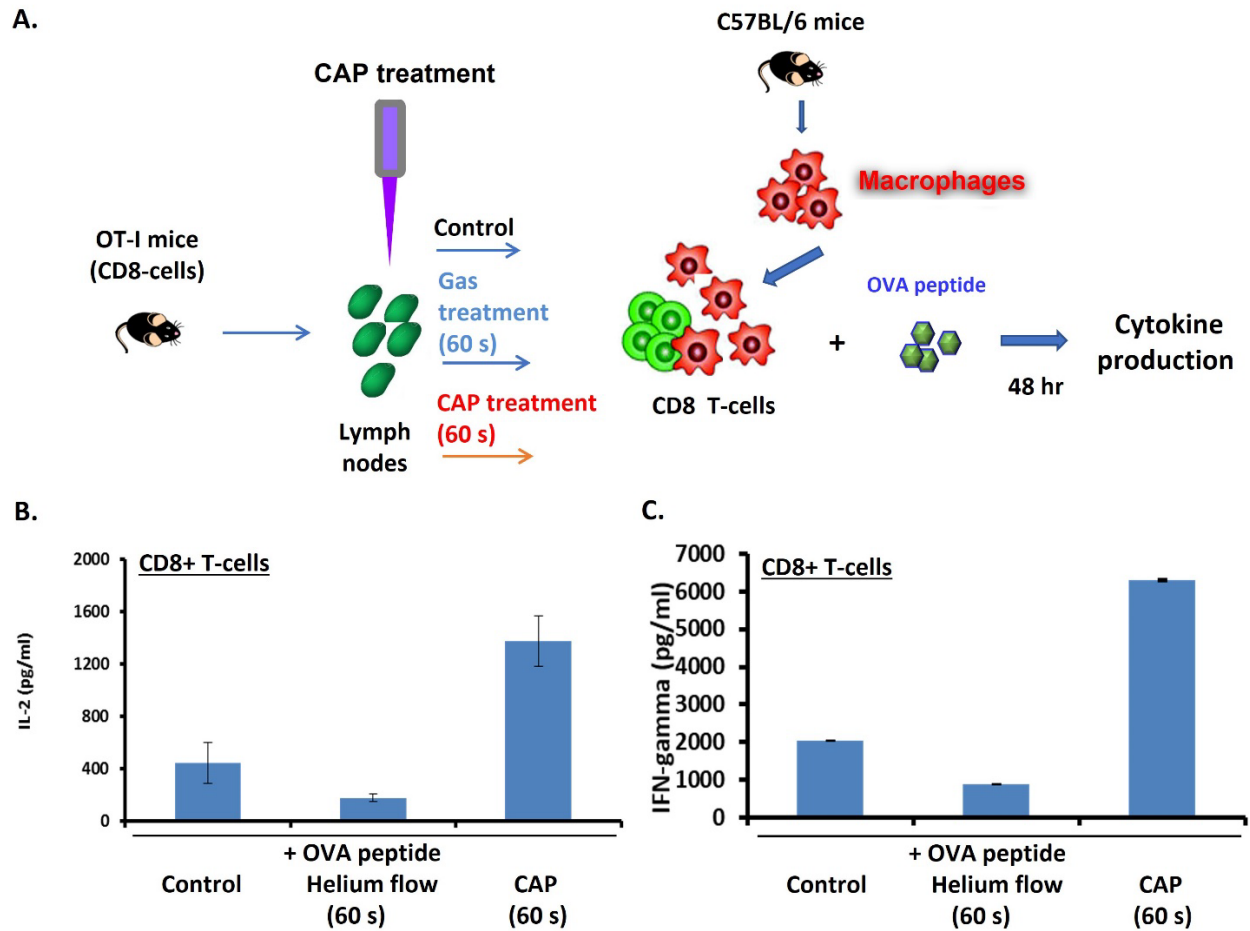

**Figure 5. Increased effector function of T-cells isolated from lymph nodes exposed to CAP *ex-vivo*. (A)** Experimental design. **(B)** IL-2 or IFN-γ production were determined by ELISA. (n=3).

**Adoptively transferred CAP-treated T-cells delayed tumor growth *in vivo*.**

The field of T-cell adoptive immunotherapy has gained particular attention in recent years given the impressive clinical responses and potential cures achieved with chimeric antigen receptor (CAR) T-cells in the treatment of several B-cell malignancies. However, not all cancer patients benefit with these therapies and as such, novel technologies that are easy to deliver and able of enhancing antitumor immune responses are greatly needed. To determine whether CAP could represent such a novel treatment approach, we exposed T-cells isolated from OT-I mice to CAP treatment, helium gas treatment, or no exposure as the control group *in vitro*. Then, CD8+T-cells were stimulated *in vitro*, followed by their adoptive transfer into tumor bearing mice **(Fig. 6A).** As shown in **Fig. 6B**, adoptively transferred CAP-treated CD8+ T-cells significantly delayed the growth of EL4 tumor *in vivo* as compared to tumor bearing mice adoptively transferred with either CD8+ T-cells exposed to helium or no exposure (control T-cells).

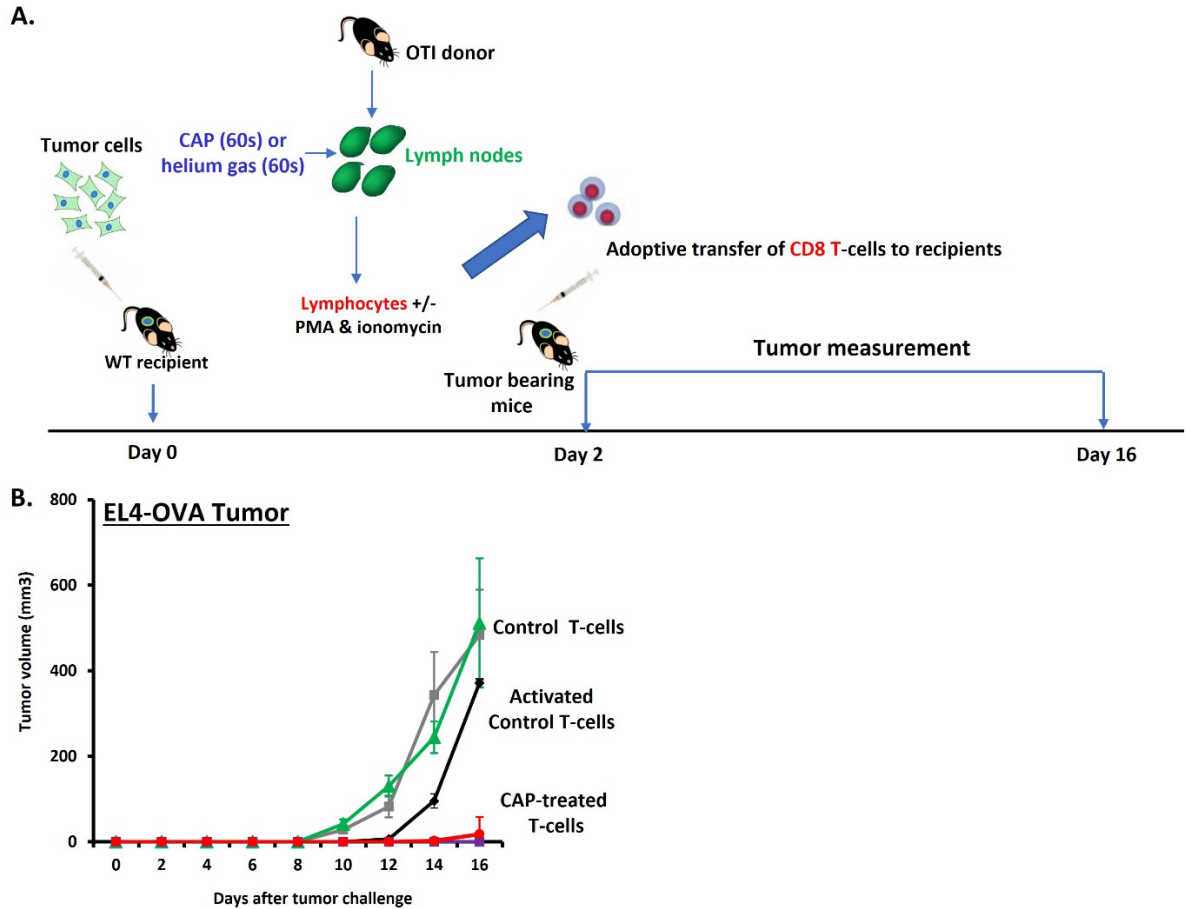

**Figure 6: Adoptive transfer of CAP-treated T-cells resulted in delayed tumor growth *in vivo***
(A) Experimental Design. (B) C57BL/6 mice were injected with tumor cells. Five tumor-bearing mice per group were evaluated. Tumor growth was assessed by using a caliper.

**Discussion.**

Immune activating and pro-inflammatory effects of CAP have long been recognized in wound treatment studies involving CAP [28]. With wound healing being the most historical and clinically utilized CAP application and immune activity playing a major role in wound healing, the relevant studies may provide a wealth of information into the immune activating capacity of CAP. The primary focus on immune activation during wound healing has been within the inflammatory phase of the wound healing process, where pro-inflammatory cytokines are secreted in order to recruit immune cells to the injury site to eradicate pathogenic microbes [28]. CAP treatment of fibroblasts has been shown to induce expression of pro-inflammatory cytokines IL-6, IL-8, MCP-1 or TGF-ß1/2, and the chemokine CXCL-1 [29].These molecules are typically expressed by fibroblasts, keratinocytes, and macrophages immediately following cutaneous injury, recruiting active inflammatory cells to the injury site. Besides, CAP treatment has been demonstrated to amplify this natural response by over-expressing these pro-inflammatory cytokines, which is now suggested to be responsible for the improved and accelerated immune defense during the inflammatory phase of wound healing [29].

Currently emerging biomedical applications of CAP also show promising results in the cancer therapy. Advancements in cancer immune therapies and immuno-oncology have provided new avenues by which to utilize immune cells against the numerous hard to treat cancers. CAP owns a great advantage to effectively deliver its reactive components to tumorous tissues and trigger strong killing effect by either direct cellular destruction or indirect immune cell death due to the activation of immune systems in the tumorous microenvironment. Here, we demonstrate how similar effects can be applied in cancer field, inducing expression of pro-inflammatory cytokines to initiate clearance of tumor cells following CAP treatment. We have shown increased expression of pro-inflammatory cytokines IL-6 and IL-12, and decreased expression of immunosuppressive molecule PD-L1 in peritoneal macrophages and T-cells treated with CAP *ex vitro*. Typically, this cytokine profile corresponds to a heightened pro-inflammatory state, similar to what would be expected upon these immune cells encountering a foreign pathogen. While the only pro-inflammatory stimulation in this study was CAP treatment, we observed a similar cytokine profile to traditional lipopolysaccharide (LPS) stimulation. More thorough investigations are necessary to determine any potentially shared mechanisms of action. Furthermore, these pro-inflammatory phenotypes enhanced antigen-presentation and effector cell function, suggesting potential to utilize this process in improving pathogen recognition and subsequent target engagement by immune cells.

With a clinical goal of utilizing this phenomenon in an immune-oncology setting, *ex-vivo* CAP treated CD8 T-cells were adoptively transferred to tumor bearing mice, significantly slowing the growth of EL4 tumors *in vivo*. Moreover, these results showed a similarly decreased tumor growth rate as is seen with conventional chemotherapy agent. A population of CD8 T cells with stem-like surface markers have now been associated with the effectiveness of tumor cell killing in adoptive cell transfer, positive human patient responses to adoptive cell transfer, and are thought to mediate the tumor cell killing process. While stem-like surface marker expression was not in the scope of this study, the high degree of deceleration of tumor growth we observed begs the question of the involvement of these stem-like CD8+ T cells and any activating effect that *ex-vivo* CAP treatment may confer. Together, these findings further demonstrate CAP's potential in the field of immunotherapy as a non-toxic and quickly delivered technology, capable of augmenting or enhancing pro-inflammatory phenotypes and downstream antitumor activity.

**Conclusion.**

In this study, we demonstrated the promising translational application of a CAP jet in cancer immunotherapy. CAP enhanced the rat peritoneal exudate macrophages' function *in vitro* by triggering an inflammatory phenotype, including the decreased expression of the immunosuppressive molecule PD-L1 and the significantly increased level of pro-inflammatory cytokines IL-6 and IL-12. Additionally, the CAP-treated macrophages triggered a better antigen-presentation to the isolated antigen-specific CD4+T-cells from the spleen of OT-II mice by producing higher level of IL-2 and IFN-γ. CAP treatment also improved cytokine production (IL-2, IFN-γ) in the T-cells isolated from the lymph nodes of C57BL/6 mice, which may be critical in orchestrating effective antitumor immune responses. Furthermore, CAP treatment changed the microenvironment in lymph nodes to affect T-cells' function. We observed the improved cytokines (IL-2, IFN-γ) of the isolated T-cells from an *ex-vivo* CAP-treated isolated lymph nodes from OT-I mice. Finally, the CAP-stimulated T-cells (isolated from OT-1 mice) were adoptively transferred into tumor bearing mice, which significantly delayed the growth of EL4 tumor *in vivo* compared with control group. In summary, CAP is a novel, safe, and effectively deliver tool to enhance the immune cells' function particularly when it is used a component of T-cells adoptive immunotherapy strategies.

**Author contributions:** F.C. and D.Y. planned and performed the experiments and help with writing the manuscript. JC and ZW performed experiments and help with reviewing the manuscript. EMS and MK supervised the experimental design and progress of this research project and wrote the manuscript.

**Acknowledgements:** The authors would like to acknowledge the flow cytometry core facilities at George Washington University for their technical support for this research project. The authors gratitude extends to the animal research facilities at George Washington University for their support. This work was supported by National Science Foundation grant, grant number 1747760.

**Conflict of Interest Disclosure:** The authors declare no interest conflict.